# Lightweight design and analysis of optical cover plate for exoplanet imaging coronagraph

Lingyi Kong[a,b], Mingming Xu[*,a,b], Wei Guo[a,b], Jiangpei Dou[a,b], Bo Chen[a,b] and Shu Jiang[a,b]
[a] Nanjing Institute of Astronomical Optics & Technology, Chinese Academy of Sciences, Nanjing 210042, China;
[b] CAS Key Laboratory of Astronomical Optics & Technology, Nanjing Institute of Astronomical Optics & Technology, Nanjing 210042,China

**ABSTRACT**

In order to reduce the load mass and solve the problem that the aluminum alloy optical cover plate of exoplanet imaging coronagraph was easy to deform, based on the equal generation design method, this paper designed and determined the configuration of the carbon fiber optical cover plate. Through the simulation of layup by finite element analysis, this paper researched the influence of different layering angles and sequences on the stiffness of optical cover plate. Finally, the carbon fiber layup method was determined as [15/-75/-15/75]s. The dynamic response analysis show that all the indexes satisfy the system requirements, and verify the feasibility of carbon fiber optical cover plate.

## 1. INTRODUCTION

The Chinese Survey Space Telescope (CSST) represents the first large-scale space survey telescope independently developed by China. It aims to conduct pioneering scientific research on the formation and evolution of cosmic structures, dark matter, dark energy, exoplanets, and solar system bodies. Notably, the exoplanet imaging coronagraph is a critical observational instrument within CSST, tasked with high-contrast imaging detection of exoplanets[1]. This capability holds significant scientific importance for China's autonomous exploration of exoplanets and the quest for a "second Earth." The exoplanet imaging star coronagraph functions as an internal occulting star coronagraph that can effectively mask or eliminate starlight by adjusting the optical path cover structure within the instrument[2]. As part of its optical components, the optical cover plays essential roles in suppressing stray light, assisting thermal control, and protecting other optical elements[3]. While its significance may not rival that of the mirror chamber assembly or camera assembly, lightweight design is imperative due to its complex structure, high processing difficulty, and substantial mass proportion attributed to aluminum alloy materials; these factors directly influence both quality and dynamic characteristics of the optical component[4]. The main supporting structure products of HiRISE high resolution camera, such as main bearing plate, truss rod and hood, are made of carbon fiber composite materials.[5]

Currently employed structural materials for space applications include aluminum alloys, titanium alloys, carbon fiber composites, and high-volume aluminum-based silicon carbide composites[6]. Yang Shuai et al. designed the M40J carbon fiber composite camera frame for a light and small space camera, which was optimized by simulation and verified by the ground mechanical environment test of the whole machine[7] . Among these materials, carbon fiber composites offer advantages such as low density, high specific stiffness, minimal thermal distortion along with linear expansion coefficients conducive to structural design flexibility; thus they are extensively utilized in space cameras, satellites, and telescopes[8]. For instance, the Hubble Space Telescope employs a trussed carbon fiber support structure connecting primary and secondary mirrors while exhibiting exceptional stability under various physical influences including heat and force interactions. The James Webb Space Telescope's optical components and integrated scientific instrument modules are constructed from M60J carbon fiber—the highest modulus PAN-based carbon fiber available today. Chinese remote sensing satellite "Jilin-1" has successfully transmitted clear remote sensing images utilizing a combination of high-precision thin-wall cylinders alongside truss rod-supported primary/secondary mirror structures made from carbon fiber composite materials. Qi Guang et al. Designed a hood composed of carbon fiber/epoxy composite material for an off-axis tri-mirror optical remote sensor. Analyses regarding stray light performance, machine tests, together with thermal vacuum and thermo physical evaluations demonstrate that this hood exhibits commendable stability, reliability, and meets stringent requirements for space utilization.

* Corresponding Author, Mingming Xu, No.188 Bancang Street, Xuanwu District, Nanjing, Jiangsu Province, China; E-mail: mingxu@niaot.ac.cn.

In this paper, T700 carbon fiber composite is used as a substitute material for aluminum alloy optical cover plate to reduce the weight of optical cover plate. Based on finite element analysis and equal substitution method, the influence of layering on the stiffness and dynamic response of the optical cover plate was studied, and the layering sequence of the optical cover plate was confirmed, providing a reference for the optimal design and processing of the optical cover plate.

## 2. OPTICAL COVER PLATE STRUCTURE

### 2.1. Material selection

Comprehensive material properties (elastic modulus, specific stiffness and thermal stability), economy, processing technology and other characteristics, in order to fully ensure the structural stability requirements, this paper uses T700 as the optical cover plate structural material, the optical cover plate of this material is 1.68kg, only 48.5% of the quality of aluminum alloy optical cover plate. Table 1 shows the performance parameters of the T700.

Table 1. Parameters of carbon fiber composites

| Attribute type | characteristic value |
|---|---|
| Longitudinal stiffness $E_x$ /GPa | 115 |
| Lateral stiffness $E_y$ /GPa | 6.43 |
| Out-of-plane stiffness $E_z$/GPa | 6.43 |
| Poisson's ratio $\upsilon_{yz}$ | 0.34 |
| Poisson's ratio $\upsilon_{xy},\ \upsilon_{xz}$ | 0.28 |
| Shear modulus $G_{xy},\ G_{yz},\ G_{xz}$/GPa | 5.5 |
| Shear modulus /GPa | 5.3 |
| Density $\rho$ /( kg /m3 ) | 1800 |

### 2.2. Basic configuration of optical cover plate

Due to the complex optical system design, compact structure and small space envelope size, the optical cover plate adopts a split thin-wall structure to protect the optical system and provide an external profile for thermal control cladding. The optical cover plate is installed on the upper part of the optical substrate, and screws are used to connect the front and back of the optical substrate and between the optical substrate and the optical substrate. The inner side of the optical cover plate is equipped with reinforcing ribs to improve the compression and shear resistance of the cover plate, reduce the bending deflection and reduce the risk of instability. The structure of the optical cover plate is shown in Figure 1.

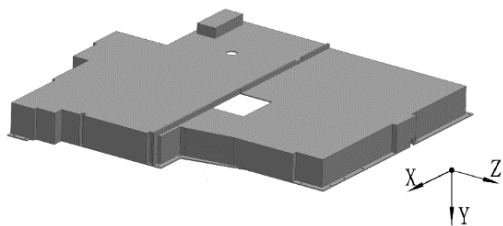

Figure 1. Model of structure

Since the optical cover plate is a structural part, in order to avoid resonance with the rocket during launch, the fundamental frequency of the optical cover plate is required to be greater than 100Hz. At the same time, the optical cover plate can withstand the dynamic load conditions in the launch stage, the specific conditions are shown in Table 2. Under this condition, the strength of the optical cover plate should meet the requirements, and the displacement response value should be less than 0.5mm to avoid collision with other parts.

Table 2. Condition of vibration

| Vibration type | direction | vibration condition | | | |
|---|---|---|---|---|---|
| Random vibration | X | Frequency Hz | 10~20 | 20~200 | 200~2000 |
| | | magnitude | 9dB/oct | 0.075$g^2$/Hz | -3dB/oct |
| | | RMS：6.96g | | | |
| | Y、Z | Frequency Hz | 10~20 | 20~200 | 200~2000 |
| | | magnitude | 9dB/oct | 0.0315$g^2$/Hz | -3dB/oct |
| | | RMS：4.5g | | | |
| Sinusoidal vibration | X | Frequency Hz | 10~15 | 15~30 | 30~100 |
| | | magnitude | 8.83mm | 8g | 5.5g |
| | Y、Z | Frequency Hz | 5~15 | 15~70 | 70~100 |
| | | magnitude | 6.62mm | 6g | 4g |

**2.3. Structure paving scheme**

According to the mechanical theory of composite laminates, the single layer thickness, lay-up Angle and lay-up sequence affect the tensile stiffness matrix and bending stiffness matrix of the laminates, and then affect the internal forces and moments of the laminates in all directions. In the lay-up design of laminate, orthogonal balanced symmetrical lay-up should be adopted as far as possible to avoid the influence of tension-shear and tension-bending coupling effects. In order to study the influence of layup direction on the performance of optical cover plate, this paper selects the orthogonal layup scheme from 0° to 90° at an interval of 15°. For details, see Table 2, where 0° is the X direction in Figure 1.

Table 3. Layering scheme

| Number | Layering scheme |
|---|---|
| 1 | [0/90/0/90]s |
| 2 | [15/-75/-15/75]s |
| 3 | [30/-60/-30/60]s |
| 4 | [45/-45/45/-45]s |
| 5 | [0/90/45/-45]s |
| 6 | [0/-45/90/45]s |

## 3. FINITE ELEMENT ANALYSIS

The optical cover plate is a thin-walled structure, so the carbon fiber composite optical cover plate adopts Shell elements, with a total of 51563 shell elements and 53123 nodes. The corresponding finite element model is shown in Figure 2.

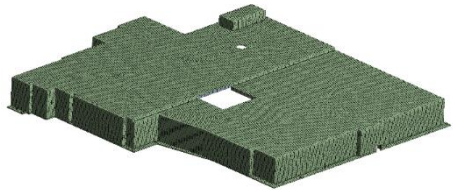

Figure 2. FEM of Carbon fiber

**3.1. Stiffness check**

"The natural frequency of the optical cover plate represents an intrinsic characteristic of its structural design, which is directly correlated with the stiffness of the structure and serves as a critical parameter in the design of aerospace load-bearing structures. Stiffness verification is conducted through modal analysis, articulated in terms of natural frequencies, corresponding modes, and mode participation coefficients; these elements form the foundation for sinusoidal frequency response analysis and random vibration assessment.

The fundamental finite element equation governing modal analysis is expressed as follows:

$$[M]\{\ddot{u}\} + [K]\{u\} = 0 \quad (1)$$

where [M] denotes the mass matrix of the cover plate; $\{\ddot{u}\}$ signifies the generalized acceleration vector at each node; [K] represents the elastic stiffness matrix of the cover plate; and $\{u\}$ indicates the generalized displacement vector at each node.

The mounting holes located at the base of the cover plate are fixed, facilitating modal solutions and analyses for each layer's cover plates listed in Table 2. The pertinent results are presented in Table 4, while Figure 3 illustrates the first six vibration modes associated with Scheme 2."

Table 4. Natural frequency

| Solution | order 1/Hz | order 2/Hz | order 3/Hz | order 4/Hz | order 5/Hz | order 6/Hz |
|---|---|---|---|---|---|---|
| 1 | 113.5 | 172.5 | 233.6 | 267.5 | 293.1 | 337.6 |
| 2 | 116.5 | 167.7 | 232 | 269.5 | 290.5 | 341.5 |
| 3 | 105.9 | 153 | 210 | 244.9 | 272 | 319.9 |
| 4 | 96.5 | 147 | 196 | 227.8 | 267 | 306.2 |
| 5 | 114.6 | 156.6 | 221.9 | 261.7 | 281.7 | 344.1 |
| 6 | 113.6 | 157 | 222.8 | 258.8 | 278.8 | 338.9 |
| **Aluminium alloy** | 117.8 | 167.1 | 227.4 | 269.8 | 308.6 | 361.2 |

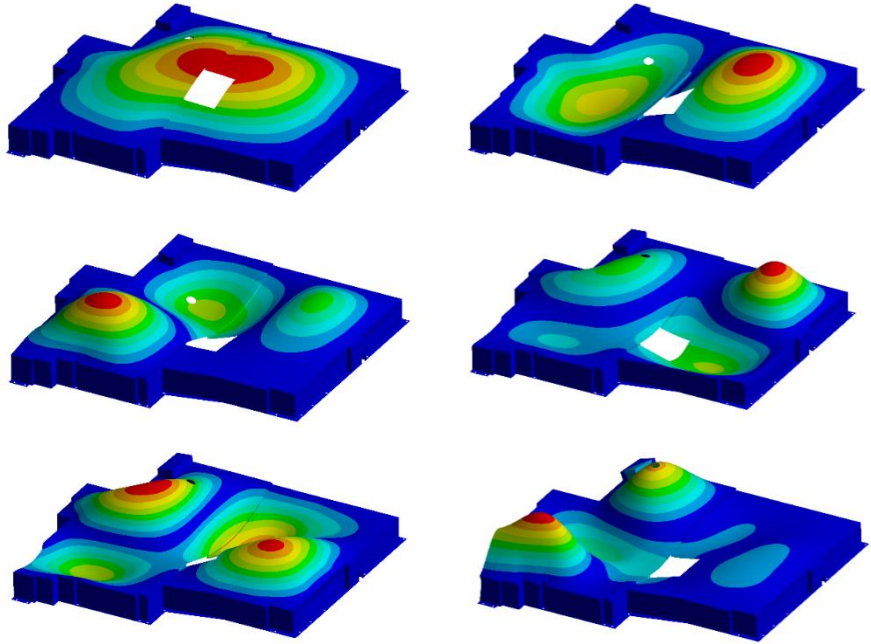

Figure 3. Mode shapes maps of optical cover plate

It can be seen from the first six modes that the response of the cover plate is normal, and its normal stiffness is relatively weak. At the same time, the cover plate is a shell structure with the worst rigidity in the middle. Therefore, under normal load, the central part responds the most. In the fiber direction of 0°~90° layering, 0° and 90° have good tensile and compressive properties, and ±45° can improve the shear resistance, while the normal stiffness of the cover plate is weak. Therefore, increasing the bidirectional component of 0° and 90° in the layering can effectively improve the tensile and compressive strength of the cover plate, and then improve the overall stiffness. With the increase of fiber Angle, the first-order natural frequency of the cover plate increases first and then decreases. According to the results of several schemes, scheme 2 has better overall stiffness and can meet the requirement of fundamental frequency greater than 100Hz. Therefore, a further analysis is carried out on the cover plate of Scheme 2.

## 3.2. Random vibration analysis

Since the cover plate is made of carbon fiber composite material, its damping coefficient is 0.03. With reference to the given conditions in Table 2, random vibration analysis of the cover plate in three directions XYZ is carried out respectively, and the analysis results are as follows.

Table 5. Response results under random vibration

| Node Serial number | Node position | RMS (g) | | |
|---|---|---|---|---|
| | | X | Y | Z |
| 1 | Top of filter wheel | 6.9622 | 4.5211 | 4.5057 |
| 2 | Clearhole | 6.961 | 12.967 | 4.509 |
| 3 | Optical cover center | 6.95999 | 16.217 | 4.5103 |

It can be seen from the response results that the response of the cover plate in the X and Z directions has no amplification, while the response in the Y direction is the most obvious, and the amplification factor is about 3.6, which is consistent with the direction of the cover plate vibration mode. Output the displacement and stress response of the cover plate in the Y direction. The maximum displacement response is 0.286mm, located in the middle of the cover plate, which is better than the required 0.5mm, and the displacement response is consistent with the first-order mode. The maximum stress response is 20.132MPa, much lower than the T700's tensile strength limit of 700MPa.

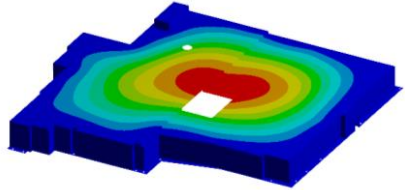

Figure 4. Y-direction displacement cloud map under random vibration

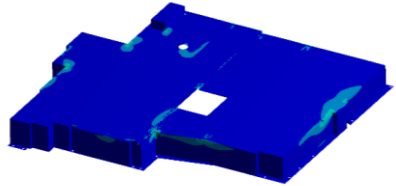

Figure 5. Y-direction stress cloud map under random vibration

## 3.3. Sinusoidal vibration analysis

In addition to random vibration, the cover plate will also be subjected to sinusoidal periodic vibration load. For the structure, the damage modes are low frequency displacement failure and high frequency acceleration failure. The low-frequency displacement amplitude in Table 2 is converted into acceleration amplitude according to formula (2), and the excitation curve is shown in Figure 6.

$$a = \frac{df^2}{250}g = 0 \tag{2}$$

Where $d$ is the displacement amplitude and $f$ is the frequency.

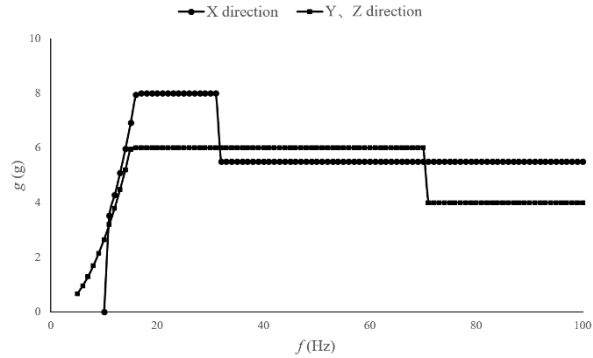

Figure 6. Sinusoidal excitation curve

The modal superposition method was used for harmonic response analysis, damping coefficient was taken as 0.03, excitation curves in three directions XYZ were input, and the output cover plate response was shown in Table 6.

Table 6. Response results under sinusoidal vibration

| Node Serial number | Node Position | Acceleration (g) | | |
|---|---|---|---|---|
| | | X | Y | Z |
| 1 | Top of filter wheel | 7.9e-4 | 0.833 | 1.58e-3 |
| 2 | clearhole | 2.2e-4 | 9.42 | 2.28e-3 |
| 3 | Optical cover center | 6.3e-4 | 17.17 | 1.88e-3 |

According to the sinusoidal vibration response results, the response of the cover plate is very small in the X and Z directions, and the maximum response in the Y direction is 17.17g with a magnification of 4.3 times, which is located in the middle of the cover plate, consistent with the modal shape and random vibration response results. The maximum displacement response of the cover plate is 0.42mm < 0.5mm, and the maximum stress is 29.78MPa, which is lower than the tensile strength limit of T700 and basically consistent with the random vibration stress distribution.

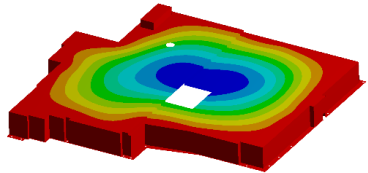

Figure 7. Y-direction displacement cloud map under sinusoidal vibration

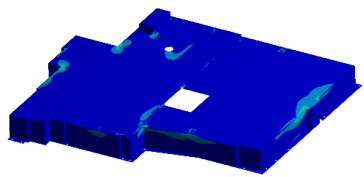

Figure 8. Y-direction stress cloud map under sinusoidal vibration

## 4. CONCLUSION

In this paper, a split carbon fiber optical cover plate is designed to meet the lightweight requirements of exoplanet imaging coronagraph. The stiffness and dynamic finite element analysis show that the optical cover plate modal and dynamic response meet the design requirements and can withstand the vibration environment requirements during launch. At the same time, the weight of the aluminum alloy structure is reduced by more than 50%, which effectively reduces the quality characteristics of the whole machine, and can guide the subsequent processing and manufacturing of the optical cover plate.